\documentclass[aps,prb,twocolumn,groupedaddress,longbibliography]{revtex4-2}%
\usepackage{epsfig,color}
\usepackage{amsmath}
\usepackage{amsfonts}
\usepackage{amssymb}
\usepackage{graphicx}
\usepackage{soul}
\usepackage{xcolor}
\newcommand\kb[1]{\textcolor{blue}{#1}}

\newcommand\red[1]{\textcolor{black}{#1}}
\begin{document}
\title{Origin of the gossamer ferromagnetism in MnTe}
\author{I. I. Mazin}
\affiliation{Department of Physics and Astronomy, George Mason University, Fairfax, VA
22030, USA}
\affiliation{Quantum Science and Engineering Center, George Mason University, Fairfax, VA
22030, USA}

\author{K. D. Belashchenko}
\affiliation{Department of Physics and Astronomy and Nebraska Center for Materials and Nanoscience, University of Nebraska-Lincoln, Lincoln, Nebraska 68588, USA}

\begin{abstract}
Absence of net magnetization in altermagnets is both a blessing (no stray fields) and a curse (no obvious way to manipulate altermagnetic domains by external fields). Yet, MnTe was demonstrated experimentally to have no measurable stray fields and yet controllable by external magnetic field --- a win-win situation. In this paper we discuss possible mechanisms driving this ultra-small canting of Mn moments, and, most importantly, the microscopic mechanism of coupling the canting with the altermagnetic order. It appear to be a higher (third) order effect in (already very small) spin-orbit coupling, which explains the unusually weak, barely measurable ferromagnetism in MnTe. Microscopic understanding of the beneficial properties of MnTe opens a road to controllable design of similar altermagnets for spintronics applications.

\end{abstract}
\maketitle
\section{Introduction}
Altermagnetism has attracted a lot of attention recently\cite{PRX,perspective,Punch}. Both novel physical phenomena and potential applications with orders of magnitude better performance have been discussed. Already more than a hundred of altermagnets have been identifies among existing antiferromagnets or predicted theoretically.

Nevertheless, nearly all experimental studies so far have been performed on a handful of materials. Till recently, the most popular was RuO$_2$, where in 2017 weak antiferromagnetism (Ru moments $\sim 0.05\ \mu_B$) was reported in neutron scattering experiment\cite{berlijn}, with an altermagnetic ordering pattern. Close to 25 papers have been published in the last few years claiming various altermagnetic manifestations. 
Unfortunately, in most cases an interpretation in terms of altermagnetism required an order-of-magnitude larger moment than that reported in Ref. \cite{berlijn}, and, to add insult to injury, it was recently shown that even the reported moment was an artifact resulting from multiple neutron scattering and in reality RuO$_2$ is nonmagnetic both in bulk and in films\cite{uSR,uSR2}, albeit may be magnetic in a few atomic layer form\cite{SHG}.

The second in popularity altermagnetic candidate is Mn$_5$Si$_3$. There the problem is different. The materials is well studied experimentally, and three ordered magnetic phases exist in the bulk\cite{Mn5Si3-1992,Mn5Si3-1995,Mn5Si3-2012}, one collinear and two noncollinear. However, all three order with the vector \{0,1/2,0\}, so by definition 
not altermagnetic. An argument was put forward that in thin films the order can switch spontaneously to \{0,0,0\}, but the fact that three different magnetic patterns share the same ordering vector suggests existence of strong antiferromagnetic coupling between the neighboring cells along the $b$ direction unit cells, which would be difficult to overcome in thin films.

This leaves us with the third by popularity compound, MnTe. This is truly a poster child for the altermagnetic case. Its magnetic structure is perfectly well known, and is definitely antiferromagnetic. It is self doped (hole concentration $\sim 10^{18}$ cm$^{-3}$\cite{Basit}), and the N\'eel vector is in the 210 direction, which is the one compatible with anomalous transport. 
As we discuss in this paper, 
it also possesses a fortuitous combination of other parameters, making it truly an ideal candidate for both studying the altermagnetic physics and for spintronics applications.

\section{MnTe: Background}
\begin{table*}[t]
\caption{Three weak effects combining to provide altermagnetic domain control by an external field. Here $H$ is the field, $x$ is the hole concentration, $t$ characterizes one-electron hopping amplitude near the Fermi level, $J$ is a gauge of the average Heisenberg coupling, and $M$ is the Mn magnetic moment, $\approx 5\ \mu_B$. The last two line, emphasized in italics, introduce two other effects, not present in MnTe, which may however appear in other altermagnetic materials. Note that double exchange canting does not couple to altermagnetism on its own, but can couple when assisted by DMI-type interactions.}
\begin{center}%
\begin{tabular}
[c]{|l|l|c|}
\hline
effect&provides&order of magnitude\\
\hline
Feedback of spin polarization on charge distribution &symmetry lowering&SOC$^2$\\
DMI in the symmetry-lowered structure&coupling between ferromagnetism and altermagnetism&SOC\\
External field&Zeeman coupling with weak ferromagnetism&H\\
{\it Double exchange}&
{\it Weak ferromagnetism} &$Mtx/J$\\
{\it Single site anisotropy}&{\it Weak ferromagnetism} and {\it coupling to altermagnetism}&SOC$^2$\\
\hline
\end{tabular}
\end{center}
\label{tab1}
\end{table*}

MnTe, despite being a canonical Mott (or, more correctly, charge transfer)  insulator has been known to occur always in a self-doped form, 
exhibiting a sizeable conductivity and also showing anomalous Hall conductivity (AHC), discovered as early as in 1965\cite{Wasscher}.  
The fact that AHC is present in an antiferromagnetic materials was either ignored\cite{Angadi} or ascribed to weak ferromagnetism of unknown origin\cite{Wasscher}.
Recently, the interest to anomalous transport in MnTe was rekindled after researchers realized that it is altermagnetic\cite{PRX,perspective,Punch,Mazin}. AHC has been remeasured\cite{Betancourt,kluczyk2023,Chilcote,Nirmal}, and weak ferromagnetism detected and its amplitude estimated to be $\sim 2.5\times 10^{-5}$\cite{Nirmal} to $5\times 10^ {-5}$\cite{kluczyk2023} $\mu_B$/Mn, parallel to $c$. Furthermore, it has been now realized that the magnetization is too small to explain the large observed AHC, which then must have the altermagnetic origin. 

Regarding the origin of this extremely weak (``gossamer'') ferromagnetism, different hypotheses were offered. In the earliest paper\cite{Wasscher} it was pointed out that the magnetic space group (MSG) Cm'c'm, which is realized in MnTe\cite{Prague}, is, per Turov's classification\cite{turov}, compatible with ferromagnetism, while no assumptions were made about the microscopic nature of the latter.

\red{For an in-plane order parameter $\mathbf{L}$, the space group of MnTe forbids linear coupling of $\mathbf{L}$ to $\mathbf{M}$ but allows a fourth-order term \cite{Wasscher,McClarty2024} $(3L_x^2-L_y^2)L_yM_z = M_z\sin{3\phi_L}$, where $\phi_L$ is the azimuthal angle corresponding to $\mathbf{L}$. The usual bilinear Dzyaloshinskii-Moriya interaction can not generate this expression, suggesting that higher-order terms may be responsible for weak ferromagnetism.}


\red{
In several later papers\cite{Kim,Chilcote} it was suggested that excess of Mn may lead to weak ferromagnetism.}
However, a possibility of excess (interstitial) Mn or Te vacancies  seems unlikely given that in such a case one would expect electron, not hole doping. {Thus,} the question of microscopic origin of the weak ferromagnetism remains open. 

In this paper, we will demonstrate that concerted action of three separate mechanisms leads to a small magnetic canting and controlled altermagnetism. In a nutshell, there is one effect, quadratic in spin-orbit coupling, the feedback of altermagnetic ordering on the charge distribution, which marginally lowers the (charge) space group from \# 194 (P6$_3$/mmc) to \#63 (Cmcm). 
The other effect is a DMI interaction that couples the already existing weak ferromagnetic component with altermagnetism, thus allowing to manipulate altermagnetic domains with an external magnetic field. This effect only appears because of the symmetry lowering described above, and is therefore proportional to the third power of weak spin-orbit coupling. And, of course, the third effect is rather trivial: Zeeman coupling of the external field with the ferromagnetic component. This is summarized in Table \ref{tab1}.

\section{Anisotropic exchange coupling in MnTe}

\begin{figure}[th]
\includegraphics[width=.49\linewidth]{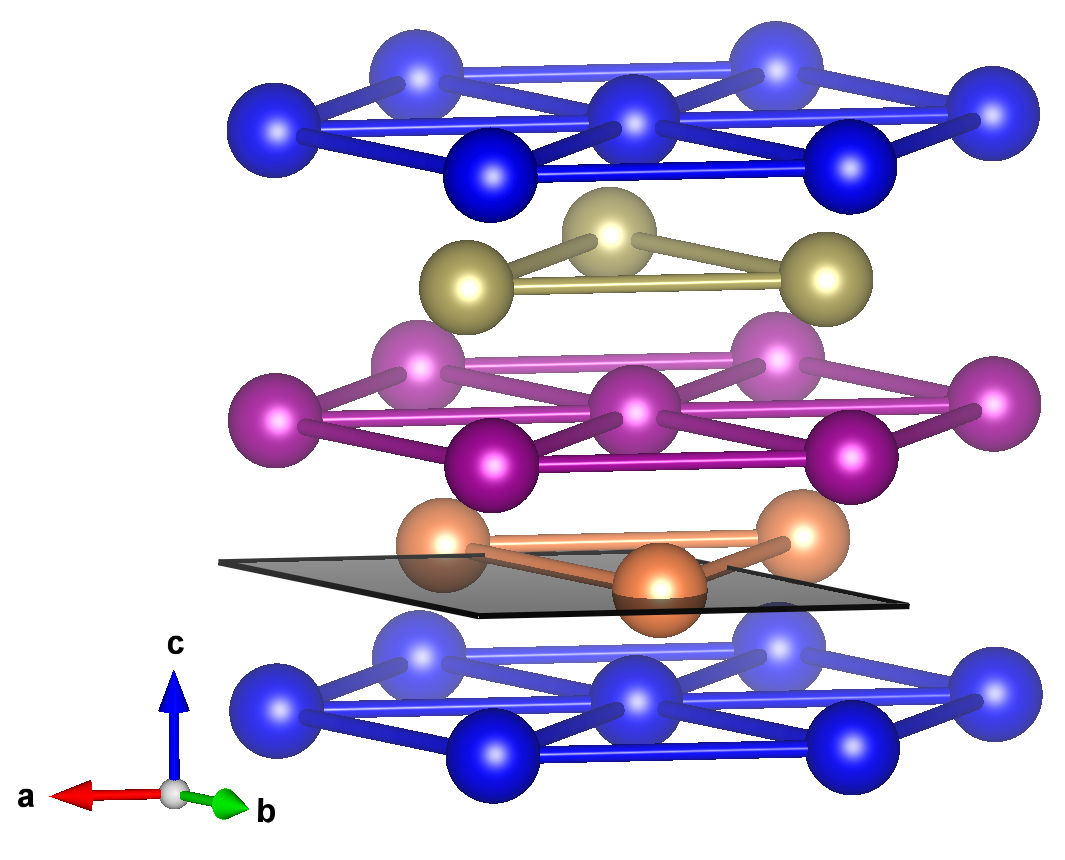}\includegraphics[width=.49\linewidth]{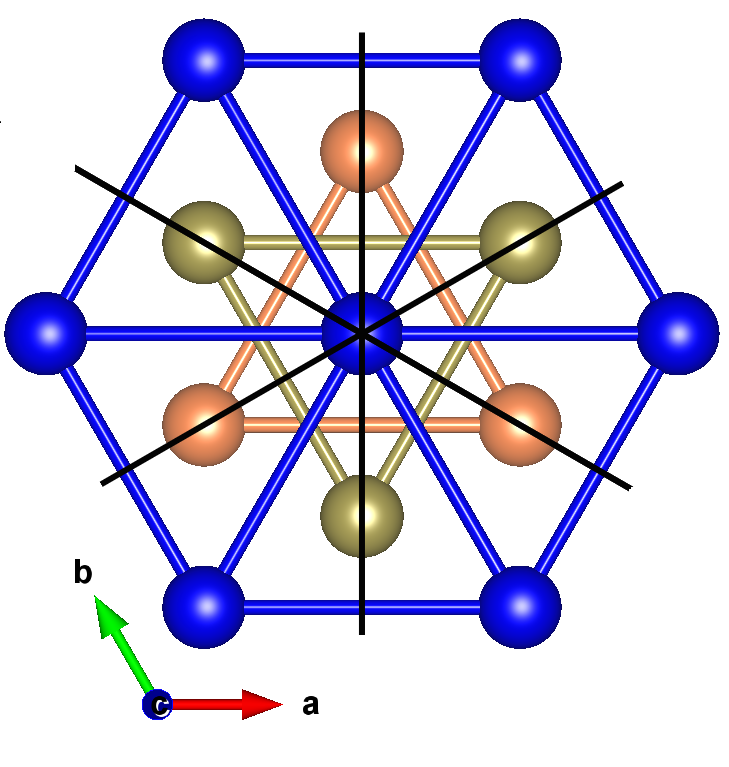}
\caption{Blue: Mn up; purple: Mn down; orange, olive: Te. Mirror planes discussed in the main text are shown in black. }
\label{str}\end{figure}\

In this Section, \red{to illustrate the lack of bilinear DMI coupling,} we will construct the full $3\times 3$ exchange matrices for the first three Mn-Mn bonds, and will show that they cannot induce a uniform canting. A full exchange matrix for a bond $I=1,2,3$ is defined such that its energy is 
\begin{equation}
    E^{(I)}=\sum_{\alpha\beta}\mathbf{m_1}\cdot \mathcal{J}\cdot\mathbf{m_2}',
\end{equation}
where $\mathbf{m}_1$, $\mathbf{m}_2$ are unit vectors in the direction of the Mn moments at the ends of the bond, and 
\begin{equation}
     \mathcal{J}=\begin{pmatrix}
J_{xx} & J_{xy} &J_{xz}\\
J_{yx} & J_{yy} & J_{yz}\\
J_{zx} & J_{zy} & J_{zz}%
\end{pmatrix}
\end{equation}
For a reference, we quote the calculated values{\cite{Mazin}} of the isotropic (Heisenberg) exchange $(J=J_{xx}+J_{yy}+J_{zz})/3$ in these three cases. The nearest neighbor bond, which is perpendicular to the basal plane, is strongly antiferromagnetic (AF), $J_1=42.1$ meV, the next nearest neighbor bond (the nearest inside the plane) has $J_2\approx 0$, and the third, which is just the sum of the other two, is also AF, $J_3=5.3$ meV.

To begin with, let us show that there is no DMI interaction on the {nearest-neighbor (nn) bond (which is AF)}. Indeed, there is a 001 mirror plane through the midpoint, which
indicates that the DMI vector $\mathbf{D}\perp\mathbf{\hat{z}}.$ The are also
three mirrors passing through the bond itself, namely 100, 010 an 110 {(Fig. \ref{str})}.
\textbf{D} must be perpendicular to each of them, which is impossible.
Moreover, a detailed analysis shows that the full exchange matrix for this
bonds is exceedingly simple:%
\begin{equation}
\mathcal{J}=%
\begin{pmatrix}
J & 0 & 0\\
0 & J & 0\\
0 & 0 & J+J_{z}%
\end{pmatrix}
\end{equation}
i.e., only the $J_{z}$ Ising exchange is allowed besides the Heisenberg exchange.

\begin{figure}[th]
\includegraphics[width=.9\linewidth]{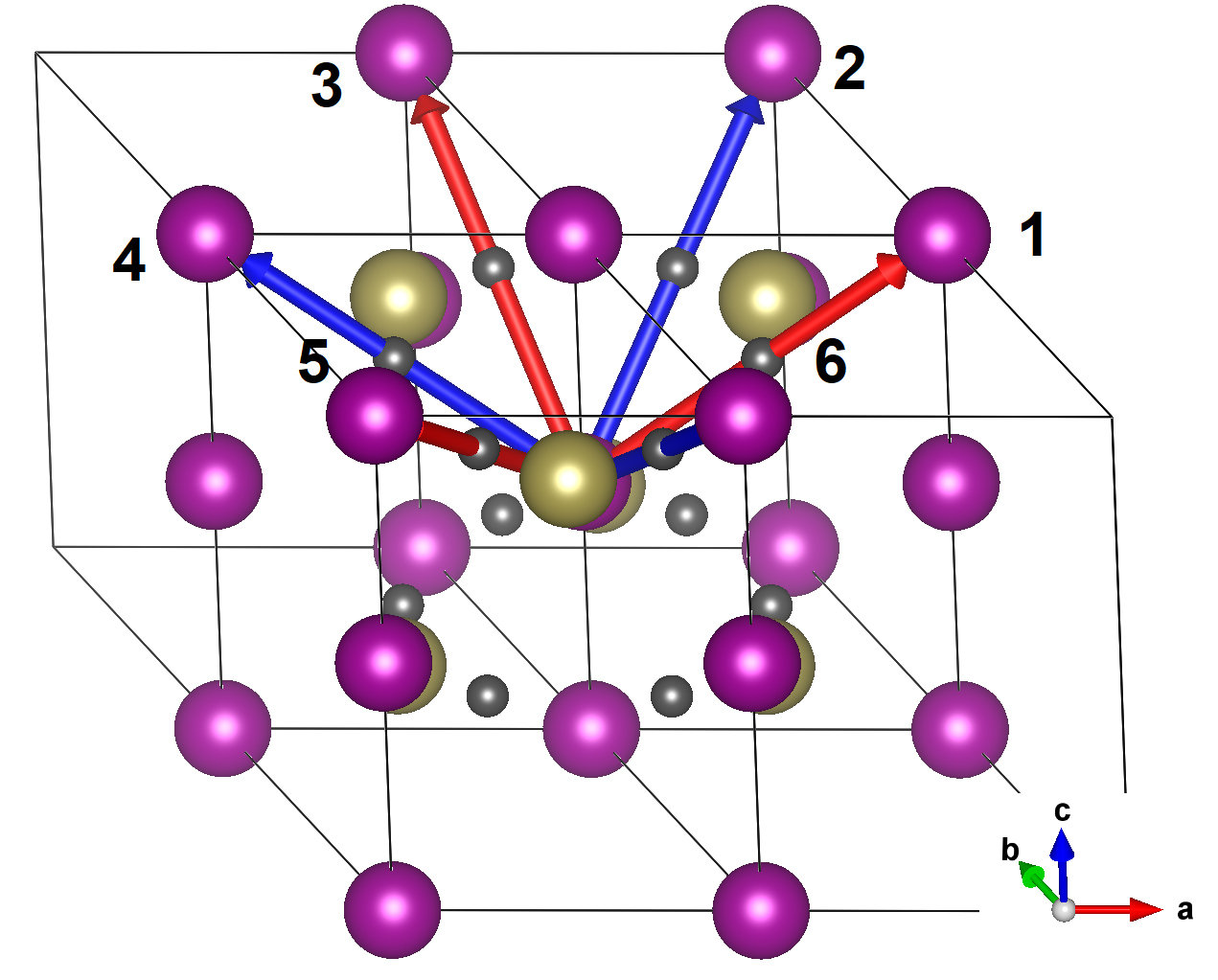}
\caption{Labeling of the Mn atoms (the upper 6) and DMI bonds (bond centers depicted with grey balls).
 The same color bonds (and their inversion partners are equivalent in the local coordinate sistem explained in the text.
 For instance, for the bond 1 the local system coincides with the global one, $x,y,z$, and for the others it
 is rotated in steps of $\pi/3$ }
\label{bonds}\end{figure}

The nn in the planes are not DMI because of inversion. Second nn across the
plane (depicted in Fig. \ref{bonds}) have a $C_{2}$ passing through the midpoint, namely
for the bond \#1, with the coordinates (1,0,$\frac{1}{2}$), the axis is 010. \textbf{D} is then perpendicular
to this axis, that is, lies in the $xz$ plane. Let us consider this
interaction more closely. There are 12 bonds of this kind, which we will label
as $i,s$, where $i=1..6$ are rotations around 001 and $s=\pm$ indicates where
the bond is going up or down. Let me also write the full exchange matrix as%
\begin{equation}
\hat{J}^{is}=%
\begin{pmatrix}
J_{XX}^{is} & J_{XY}^{is} & J_{Xz}^{is}\\
J_{YX}^{is} & J_{YY}^{is} & J_{Yz}^{is}\\
J_{zX}^{is} & J_{zY}^{is} & J_{zz}^{is}%
\end{pmatrix}
\end{equation}
where the new axis $X$ is the projection of the bond onto the $ab$ plane, and
$Y$ is perpendicular to $X,z.$ Note that is is enough to consider only $s=+,$
because due to the inversion $\hat{J}^{1-}=\hat{J}^{4+},$ $\hat{J}^{2-}%
=\hat{J}^{5+},$ $\hat{J}^{3-}=\hat{J}^{6+},$ etc. As mentioned, there is only
one operation that keeps a bond in place, $C_{2Y},$ which transforms
$S_{X}\rightarrow-S_{X,}$ $S_{Y}\rightarrow S_{Y,}$ $S_{Z}\rightarrow-S_{Z,}$
so
\begin{align}%
\begin{pmatrix}
J_{XX} & J_{XY}^{{}} & J_{Xz}^{{}}\\
J_{YX}^{{}} & J_{YY}^{{}} & J_{Yz}^{{}}\\
J_{zX}^{{}} & J_{zY}^{{}} & J_{zz}^{{}}%
\end{pmatrix}
&=%
\begin{pmatrix}
J_{XX}^{{}} & -J_{YX}^{{}} & J_{zX}^{{}}\\
-J_{XY}^{{}} & J_{YY}^{{}} & -J_{zY}^{{}}\\
J_{Xz}^{{}} & -J_{Yz}^{{}} & J_{zz}^{{}}%
\end{pmatrix}\\
&=%
\begin{pmatrix}
J_{XX}^{{}} & D_{z} & A\\
-D_{z} & J_{YY}^{{}} & D_{X}\\
A & -D_{X} & J_{zz}^{{}}%
\end{pmatrix}
,
\end{align}
therefore $J_{Xz}^{{}}=J_{zX}=A,$ $J_{XY}^{{}}=-J_{YX}^{{}}=D_{z},$
$J_{zY}^{{}}=-J_{zX}^{{}}=D_{X}.$

Next, consider a mirror plane through the origin and the Te atoms. One such
plane converts $J^{1}$ to $J^{2},$ and it also changes $S_{X}\rightarrow
-S_{X,}$ $S_{Y}\rightarrow S_{Y,}$ $S_{Z}\rightarrow-S_{Z}.$ Then%
\begin{equation}%
\begin{pmatrix}
J_{XX}^{1} & D_{z}^{1} & A^{1}\\
-D_{z}^{1} & J_{YY}^{1} & D_{X}^{1}\\
A^{1} & -D_{X}^{1} & J_{zz}^{1}%
\end{pmatrix}
=%
\begin{pmatrix}
J_{XX}^{2} & -D_{z}^{2} & A^{2}\\
D_{z}^{2} & J_{YY}^{2} & -D_{X}^{2}\\
A^{2} & D_{X}^{2} & J_{zz}^{2}%
\end{pmatrix}
,
\end{equation}

\begin{figure}[th]
\includegraphics[width=.9\linewidth]{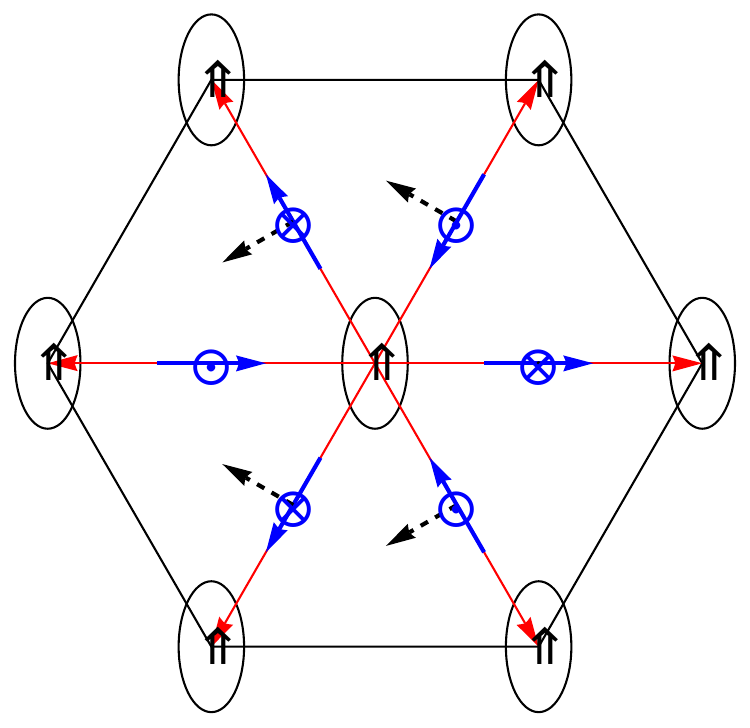}
\caption{DMI interaction. Dotted and crossed circles show $\pm D_z$, the blue arrows $D_X$, and the dashed black ones the small (tiny!) 
components that arise from the feedback of the magnetic symmetry breaking.}
\label{all}\end{figure}

That is, the compass term $A$ is the same, the $z$-component of the DMI vector
changes sign, and also the bond direction. The result is shown in Fig. \ref{all}, where
the red arrows give the local $X$ axis (the radial component of the DMI
vector, blue, is parallel). Since the altermagnetic N\'eel vector lies in the plane, the compass terms has no effect, and the DMI field exerted by the parallel to each other spin-up moments onto the the spin down subsystem cancel out. Note that this cancellation is a direct consequence of the 3-fold rotational axis.

\begin{figure}[th]
\includegraphics[width=.9\linewidth]{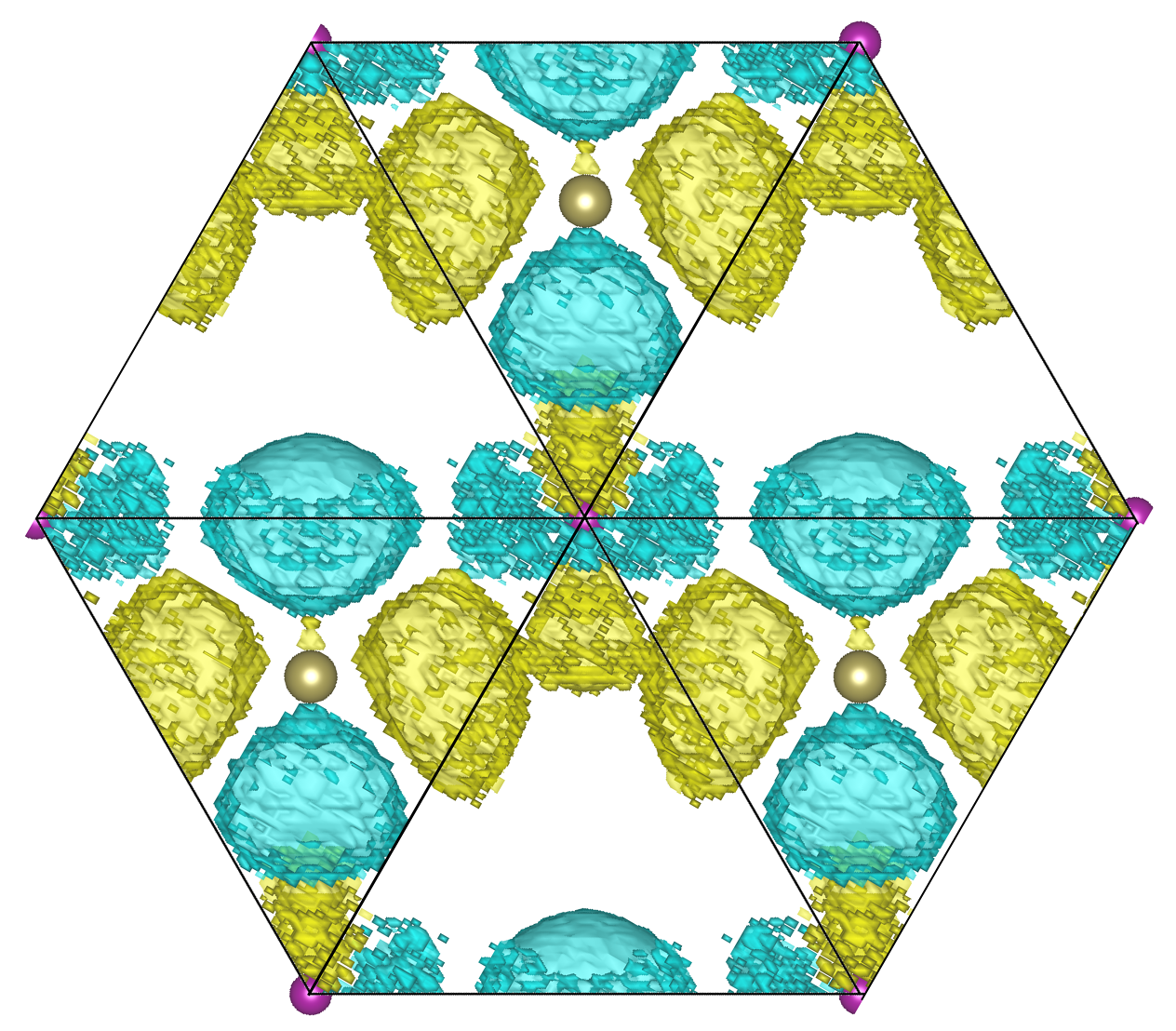}
\caption{Difference between the charge densities of the 210 N\'eel vector (the ground state) and the 100 one. The colors indicate the sign, and isosurfaces are drawn at the level of $2\times 10^{-6}$e/\AA$^3$ (about 4 orders of magnitude less than the total 
charge density). Note breaking of the $C_3$ rotational symmetry.}
\label{diff}\end{figure}
However, that does not account for the magnetic order already affecting the
crystal symmetry. As long as there is spin-orbit coupling, there will be an
additional very small change in the charge distribution ($\propto M^{2})$ that
breaks the hexagonal symmetry of the charge cloud, as shown (exaggerated) in
Fig. \ref{diff}.

In Fig. \ref{diff} we show the results of DFT calculation using the VASP code\cite{VASP1,VASP2}. Charge density was calculated self-consistently, including SOC, for the N\'eel vector $\mathbf{M}||210$ and the perpendicular to it vector $\mathbf{M}||100$. A very fine $k$-point mesh of $21  \times 21 \times  11$ was used, with the convergence criterion of $10^{-8}$ eV and an energy cut-off of 800 eV.  In order to avoid systematic error, both calculations were started with a self-consistent calculation with $\mathbf{M}||001$, and then the spin coordinate system was rotated using the SAXIS tag in VASP. The same protocol applied to nonrelativistic calculations did not produce any symmetry breaking. 

Because of this symmetry breaking, the mirror $\sigma_{Y}$ remains only for the bonds
$J^{1}$ and $J^{4},$ but is broken for the others (and, of course, $C_3$ is broken as well).
As a result, the in-plane
vector $D_{x}$ is not strictly radial any more for the other bonds, but acquires
a tiny tangential component $D_{y}.$


Let me now concentrate on the bonds $J^{2}$ and $J^{3}.$There still is a
mirror relating the two bonds. However, there is no condition any more that
$D_{Y}=0.$ Keeping only the DMI terms,%
\begin{equation}%
\begin{pmatrix}
& D_{z}^{3} & D_{Y}^{3}\\
-D_{z}^{3} &  & D_{X}^{3}\\
-D_{Y}^{3} & -D_{X}^{3} &
\end{pmatrix}
=%
\begin{pmatrix}
& -D_{z}^{2} & D_{Y}^{2}\\
D_{z}^{2} &  & -D_{X}^{2}\\
-D_{Y}^{2} & D_{X}^{2} &
\end{pmatrix}
,
\end{equation}
That is to say, there appears a small tangential component $D_{Y}.$

Now we need to connect that to $J^{5,6}.$ There is a $C_{2x}$ axis. It will
convert $J^{2}$ to $J^{6},$ and, in the original Cartesian system,
$S_{x}\rightarrow S_{x,}$ $S_{y}\rightarrow-S_{y,}$ $S_{z}\rightarrow-S_{z}.$
Without going into details, it means that the $J_{yz}$ component will not
change, that is, $D_{x}$ will remain the same, and will have a net component
after averaging, as shown in Fig. \ref{all} by the dashed arrows. 
{This net $D_x$ component, combined with the N\'eel vector parallel to $y$, will exert a net exchange field parallel to $z$ and induce a tiny net polarization $M_z$, in agreement with the experiment (albeit too small to calculate it from first principles).}
Note that, per Table 1, this interaction scaled as cube of SOC {(it is the product of the first two lines in the Table)}.

While we do not make any attempt to cast this interaction into a gauge-invariant form, from general symmetry considerations one can conclude that this interaction can be written 
generically as $H=\sum_{ij}{(\mathbf{S}_i\cdot\hat{K}\cdot\mathbf{S}_i)(\mathbf{S}_i\times\mathbf{S}_j)}\cdot \mathbf{C}_{ij}$, where the tensor $\hat{K}$ is similar to the single-site anisotropy tensor, and $\mathbf{C}_{ij}$ similar to the Dzyaloshinskii-Moriya exchange. It can be compared to the interaction 
defined by Bl\"ugel et al\cite{Stefan} and by Brinker et al\cite{Brinker} (see also Ref. \cite{Fe}) as ``chiral biquadratic interaction'', $H=\sum_{ij}{(\mathbf{S}_i\cdot\mathbf{S}_j)(\mathbf{S}_i\times\mathbf{S}_j)}\cdot \mathbf{C}_{ij}$.

\section{Other potential sources of canting}

Besides DMI, two other sources for weak ferromagnetism are often discussed in the literature. One is the single site anisotropy, \kb{which} is known\cite{NiF2} to be the source of canting in NiF$_2$. This requires easy axes that are different in  the two sublattices. For instance, in the two sublattices in NiF$_2$ these are [110] and [1$\bar{1}0$]. This mechanism also couples with the altermagnetic order (note that NiF$_2$  is an altermagnet). In MnTe, however, this mechanism is absent by symmetry: the two magnetic sublattices have exactly the same local symmetry and the same easy axes, namely [210],[120], and [1$\bar{1}0$]. 

Another mechanism commonly believed to be a possible source for weak ferromagnetism is double exchange, proposed by Zener in 1951\cite{Zener}, and by now a part of  standard textbooks on magnetism \cite{Khomskii}. In essence, it is a competition between antiferromagnetic exchange among (reasonably well) localized moments, which wants to keep the latter antiparallel, and the kinetic energy of the itinerant electrons, assuming that there is a strong interaction keeping itinerant electrons' spins locally parallel to the localized moments. This interaction can be Hund's rule coupling, $J_H$, (combined with the Hubbard $U$), or, if the itinerant electrons originate from different atomic species, Schrieffer-Wolff interaction\cite{Glasbrenner2014} (which would have been the case in the charge-transfer insulator MnTe). Conventionally, the corresponding Hamiltonian is written under the assumptions that (a) the undoped material is insulating, (b) the Hund's rule coupling is infinite, so electron hopping along an antiferromagnetic bond is fully arrested and (c) all nearest neighbor bonds are antiferromagnetic and hopping to all other neighbors is neglected. The model is usually illustrated on a 1D N\'eel antiferromagnet. In that case, in a fully collinear antiferromagnetic state the doping proceeds through a zero-width band and the kinetic energy of the doped carriers can be lowered by canting the local spins and allowing for some hopping, which in this case scales as the sine of the canting angle $\phi$, and the Hamiltonian reads \cite{Khomskii}:
\begin{equation}
E=-aJ\cos{2\phi}-bxt|\sin{\phi}|,\label{DE} 
\end{equation}
where $\phi$ is the canting angle away from the original collinear antiferromagnetic arrangement, $E$ the total energy per spin, $J$ a properly averaged antiferromagnetic exchange coupling, $x$ the concentration of itinerant carriers (assumed $x\ll 1$), $t$ the hopping parameter of the itinerant electrons, and $a$ and $b$ the geometrical coefficients characterizing the connectivity in the system. \red{Note that nonrelativistic spin-rotation symmetry requires $E(\phi)$ to be an even function; therefore, the second term in (\ref{DE}) is written as an absolute value.} Minimizing this expression, one gets 
\begin{equation}
  |\sin{\phi}|=(bt/4aJ)x, 
\end{equation}
that is to say, the net ferromagnetic moment is linearly proportional to the concentration $x$.

\red{This is the standard textbook description of the double exchange effect. However, as we discuss below, this description is incomplete. First of all, 
the assumptions above can be quantified as} $t^2/J_H\ll t|\sin{\phi}|$ and $t'\ll t|\sin{\phi}|$, where $t'$ characterizes the same-spin hopping. 
No matter what  the materials parameters are, it is clear that these conditions will be violated at sufficiently small $\phi$.

\begin{figure}[th]
\includegraphics[width=.9\linewidth]{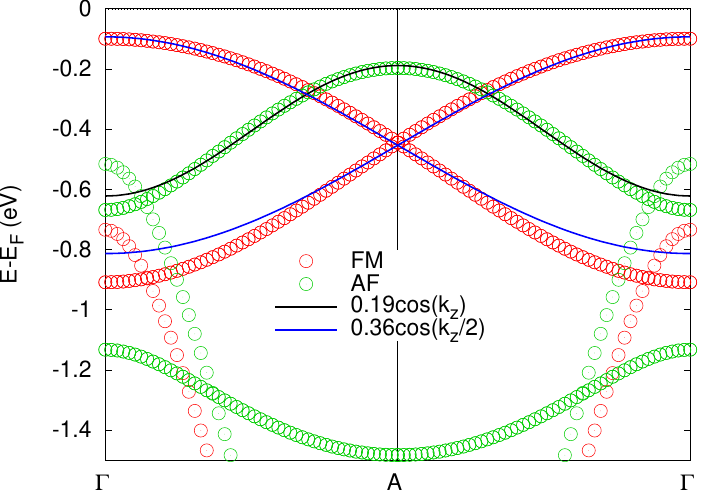}
\caption{Band dispersions for highest valence bands in ferromagnetic (FM) and altermagnetic (antiferromagnetic, AF)  states. The band dispersions near the top of the valence band along the $\Gamma$-A direction were fitted to $cos(ck_z/2$ and $cos(ck_z$, respectively. One can see that while the prefactor in the latter case is nearly twice larger, the effective net hopping along $z$ is nearly 200 meV already in the antiferromagnetic case.}
\label{disp}\end{figure}

Indeed, generally speaking, if there are several different channels for one-electron hopping, they add in quadrature, so the effective hopping amplitude $t|\sin{\phi}|$ in Eq. \ref{DE} should be replaced with $\sqrt{(t\sin{\phi})^2+ (u t^2/J_H)^2+(v t')^2}$, where $u$ and $v$ are coefficients of the order 1. The last two terms can be estimated from the band structure calculation, since they determine the band width in the antiferromagnetic case.
Since the in-plane order is collinear in both cases, the problem maps exactly onto a 1D antiferromagnetic chain, which makes this comparison particularly straightforward. Fig. \ref{disp} shows that the effective $z$-hopping $\sqrt{(u t^2/J_H)^2+(v t')^2}$ is nearly 200 meV, albeit becomes nearly twice smaller that the ferromagnetic hopping, {$\sqrt{t^2+ (u t^2/J_H)^2+(v t')^2}\approx 340$} meV, and { $t\approx 280$ meV}. Obviously, despite the seemingly favorable case (1D hopping, large onsite spin-flip energy cost), the {above conditions are far from being satisfied in MnTe in any range of canting angles}.

\red{One can also imagine a slightly different mechanism for double exchange, rarely discussed in the literature, which does not suffer from this problem: if the relevant gap-edge states, like the highest occupied states in MnTe,
are strictly spin-degenerate,
then canting 
introduces spin-splitting, $\pm t_{\mathbf{k}}\sin\phi$, which can lead  to a non-analytic energy gain that is linear in $\phi$ as long as the splitting is larger than the Fermi level. The mathematics is more complicated in this case and strongly dependent on the details of the electronic structure.}

\red{At first sight, this case might apply to MnTe, where the valence band maximum is formed by a quadruplet of states that are almost degenerate and only split by about 2 meV into two doublets at the A point \cite{Faria2023,belashchenko2024}. However, altermagnetic hybridization in MnTe maximally  entangles spin with the sublattice degree of freedom for the valence bands near the A point; as a result, any linear combination of the four valence states has a zero expectation value of the transverse spin components \cite{Faria2023,belashchenko2024}. This means that, to linear order in $\sin\theta$, the effective transverse magnetic field introduced by spin canting does not split the degeneracy at the A point, and there is, therefore, no non-analytic energy gain. Thus, the second possible mechanism of double exchange is not operative in MnTe.}

\begin{figure}[th]
\includegraphics[width=.9\linewidth]{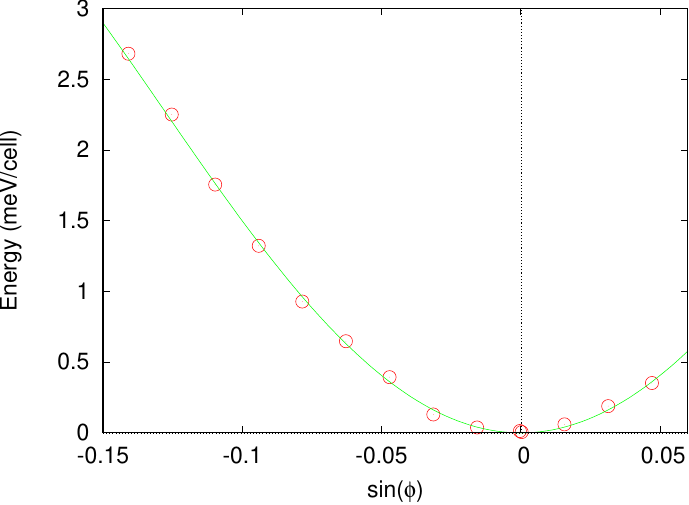}
\caption{Energy in the canted state as a function of the canting angle $\phi$, assuming hole doping of 0.2 hole/Mn. The line is a fitting to the even-only power polynomial up to the 6th power.}
\label{cant}\end{figure}

To verify this conclusion, we have calculated the total energy as a function of the weak ferromagnetic moment, using the VASP constrained-directions mode for a sizeable doping of 0.2 hole/Mn  (Fig. \ref{cant}). The calculations show no minimum at any finite canting, and a perfectly quadratic dependence for canting angles up to $\approx 3.5^\circ$ (with a slight bending down at higher angle, reflecting indeed the double exchange physics).

Suppose that an altermagnetic material is allowed to have weak ferromagnetism, but microscopically the mechanism of spin canting does not couple to altermagnetic domains, that is, canting can be flipped without incurring a change in energy even is the altermagnetic order is not flipped. This is for instance in case in easy-plane rutiles. On the first glance such canting is completely useless from the point of view of manipulating altermagnetic domains with an external field. However, if another canting mechanism is present, such as the one described in the previous section, which maybe order(s) of magnitude smaller in amplitude, than the two canting mechanisms couple to each other, and thus the former (stronger) also couples with the altermagnetic order. This is an exciting opportunity that, in principle, allows to manipulate the domains with very small fields, even if the relevant DMI interaction is of a higher order and extremely small. As discussed below, in MnTe this possibility is not realized, but it may prove instrumental in other altermagnetic materials.


 To conclude this section, while in principle there exist other mechanisms for canting not related to DMI interactions and not coupled directly to the altermagnetic order, none of them is operative in MnTe and the observed magnetization is entirely due to the chiral biquadratic interaction described in the previous section.

\section{Discussion and conclusions}
A major obstacle on the road to exciting spintronics applications is inability to manipulate altermagnetic domains by external stimuli. Indeed, the major advantage of altermagnets over ferromagnets is absence of net magnetization and thus of stray fields. Yet the same zero net magnetization prevents an altermagnet from interacting with external magnetic fields and thus presents a problem with controlling and manipulating altermagnetic domains. In many cases, in fact in those same cases where anomalous transport in an altermagnet is nonzero, the magnetic point group is compatible with ferromagnetism, a phenomenon know as weak ferromagnetism. However, several prerequisites are necessary before one can take an advantage of this weak ferromagnetic component, namely, the canting of altermagnetic moments need not only  be allowed by symmetry, but microscopic interactions must be present generating such canting. Second, this canting must be small enough not to generate stray fields by itself. Third, it has to couple with the altermagnetic order so that flipping the ferromagnetic component would lead to flipping the altermagnetic order.

Three mechanisms are known to generate weak ferromagnetism. One is a single-site anisotropy, as in NiF$_2$\cite{NiF2}, \red{which is not present in MnTe}. Another is double exchange, which doesn't couple with altermagnetic order. The third possibility is Dzyaloshinskii-Moriya interaction, but, as the discussion above exemplifies, absence of an inversion center at the midpoint of a magnetic bond does not guarantee (contrary to a common misconception) presence of such interaction, and even if it is non-zero in most cases it cancels out when summed over all equivalent bonds. Besides, if it is allowed, it is the first order in spin-orbit coupling so may actually be strong enough to generate too large stray fields.

As discussed in this paper, MnTe represents a new paradigm, applicable to altermagnetic semiconductors, doped naturally or in a controlled way by a very small number of carriers, $10^{-4}$--$10^{-5}$ per magnetic ion. This doping ensures metallic conductivity required for applications, and, in principle, albeit not in MnTe, may actually enhance the magnetization, if needed for application, via double exchange. Thus, we have three decoupled, that is, controlled by different factors, parameters: altermagnetic transport, which is completely oblivious to the small ferromagnetic component, the four-spin Dzyaloshinskii-Moriya (``chiral biquadratic'') interaction that generates weak ferromagnetism {\it and} couples the ferro- and and the altermagnetic order parameters; finally, weak ferromagnetism couples to external magnetic filed and allow for controlled manipulations of altermagnetic domains. The fact that MnTe really exists and experimentally demonstrates all these features is strongly encouraging and suggest a new road towards altermagnetic spintronics.

\section{Methods}
All calculations presented here used Vienna ab initio
Simulation Package (VASP) \cite{VASP1,VASP2} within projector augmented wave
(PAW) method.\cite{PAW1,PAW2} The Perdew-Burke-Enzerhof (PBE) \cite{GGA}
generalized gradient approximation was employed to describe exchange-correlation effects. To improve the description for localized d-electrons
in Mn$^{2+}$ ion to be strongly correlated, we added a Hubbard $U$ correction with the  fully localized limit double-counting recipe\cite{liechtenstein_density-functional_1995,dudarev_electron-energy-loss_1998}, with the effective parameter $U-J=4$ eV. Pseudopotential from the VASP library, PAW\_PBE-Te and PAW\_PBE-Mn\_pv were used, with the energy cutoff of 500 eV, and 9$\times 9\times 7$ (160 irreducible k-points) mesh.

Some of the figures were generated using VESTA software\cite{VESTA}.
\begin{acknowledgments}
 {We are} thankful to Nirmal Ghimire, Libor \^Smejkal, Jairo Sinova, and Stefan Bl\"ugel 
for valuable discussions.
IM was supported by the Army Research Office under Cooperative Agreement Number W911NF-22-2-0173,  {and KB by the U.S. Department of Energy (DOE) Established Program to Stimulate Competitive Research (EPSCoR) grant No. DE-SC0024284.}
\end{acknowledgments}
\bibliography{MnTe}
\end{document}